
\magnification=1200
\voffset=0 true mm
\hoffset=0 true in
\hsize=6.5 true in
\vsize=8.5 true in
\normalbaselineskip=13pt
\def\doublespace{\baselineskip=20pt plus 3pt\message{double space}}
\def\singlespace{\baselineskip=13pt\message{single space}}
\let\spacing=\singlespace
\parindent=1.0 true cm



\newcount\equationumber \newcount\sectionumber
\sectionumber=1 \equationumber=1
\def\setsection{\global\advance\sectionumber by1 \equationumber=1}

\def\numbe{{{\number\sectionumber}{.}\number\equationumber}
                            \global\advance\equationumber by1}
\def\numberit{\eqno{(\number\equationumber)} \global\advance\equationumber by1}

\def\numberal{(\number\equationumber)\global\advance\equationumber by1}

\def\sectionit{\eqno{(\numbe)}}

\def\ccf#1{\,\vcenter{\normalbaselines
    \ialign{\hfil$##$\hfil&&$\>\hfil ##$\hfil\crcr
      \mathstrut\crcr\noalign{\kern-\baselineskip}
      #1\crcr\mathstrut\crcr\noalign{\kern-\baselineskip}}}\,}
\def\scf#1{\,\vcenter{\baselineskip=9pt
    \ialign{\hfil$##$\hfil&&$\>\hfil ##$\hfil\crcr
      \vphantom(\crcr\noalign{\kern-\baselineskip}
      #1\crcr\mathstrut\crcr\noalign{\kern-\baselineskip}}}\,}

\def\small3j#1#2#3#4#5#6{\def\st{\scriptstyle} 
   \bigl(\scf{\st#1&\st#2&\st#3\cr
           \st#4&\st#5&\st#6\cr} \bigr)}




\def\ref#1{$^{#1)}$}


\def\upa#1{\raise 1pt\hbox{\sevenrm #1}}
\def\dna#1{\lower 1pt\hbox{\sevenrm #1}}
\def\dnb#1{\lower 2pt\hbox{$\scriptstyle #1$}}
\def\dnc#1{\lower 3pt\hbox{$\scriptstyle #1$}}
\def\upb#1{\raise 2pt\hbox{$\scriptstyle #1$}}
\def\upc#1{\raise 3pt\hbox{$\scriptstyle #1$}}
\def\hprime{\raise 2pt\hbox{$\scriptstyle \prime$}}
\def\ccom{\,\raise2pt\hbox{,}}


\def\asymptotically#1{\;\rlap{\lower 4pt\hbox to 2.0 true cm{
    \hfil\sevenrm  #1 \hfil}}
   \hbox{$\relbar\joinrel\relbar\joinrel\relbar\joinrel
     \relbar\joinrel\relbar\joinrel\longrightarrow\;$}}
\def\Asymptotically#1{\;\rlap{\lower 4pt\hbox to 3.0 true cm{
    \hfil\sevenrm  #1 \hfil}}
   \hbox{$\relbar\joinrel\relbar\joinrel\relbar\joinrel\relbar\joinrel
     \relbar\joinrel\relbar\joinrel\relbar\joinrel\relbar\joinrel
     \relbar\joinrel\relbar\joinrel\longrightarrow$\;}}

\def\dal{\mathop{\sqcup\hskip-6.4pt\sqcap}\nolimits}

\catcode`@=11
\def\C@ncel#1#2{\ooalign{$\hfil#1\mkern2mu/\hfil$\crcr$#1#2$}}
\def\gf#1{\mathrel{\mathpalette\c@ncel#1}}      
\def\Gf#1{\mathrel{\mathpalette\C@ncel#1}}      

\def\gapx{\lower 2pt \hbox{$\buildrel>\over{\scriptstyle{\sim}}$}}
\def\lapx{\lower 2pt \hbox{$\buildrel<\over{\scriptstyle{\sim}}$}}

\def\nablaleft{\hbox{\raise 6pt\rlap{{\kern-1pt$\leftarrow$}}{$\nabla$}}}
\def\nablaright{\hbox{\raise 6pt\rlap{{\kern-1pt$\rightarrow$}}{$\nabla$}}}
\def\nablaboth{\hbox{\raise 6pt\rlap{{\kern-1pt$\leftrightarrow$}}{$\nabla$}}}

\def\boks#1#2{{\hsize=#1 true cm\parindent=0pt
  {\vbox{\hrule height1pt \hbox
    {\vrule width1pt \kern3pt\raise 3pt\vbox{\kern3pt{#2}}\kern3pt
    \vrule width1pt}\hrule height1pt}}}}

\def\heading{ }
\def\range{ }

\def\body{\vfill\eject\parindent=1.0 true cm
 \ifx\spacing\singlespace\singlespace\else\doublespace\fi}
\def\title#1{\centerline{{\bf #1}}}

\def\today{\ifcase\month\or
  January\or February\or March\or April\or May\or June\or
  July\or August\or September\or October\or November\or December\fi
  \space\number\day, \number\year}
\let\date=\today
\newcount\hour \newcount\minute
\countdef\hour=56
\countdef\minute=57
\hour=\time
  \divide\hour by 60
  \minute=\time
  \count58=\hour
  \multiply\count58 by 60
  \advance\minute by -\count58

\def\sectionskip{\penalty-500\vskip24pt plus12pt minus6pt}

\def\sec{\hbox{\lower 1pt\rlap{{\sixrm S}}{\hbox{\raise 1pt\hbox{\sixrm S}}}}}
\def\section#1\par{\goodbreak\message{#1}
    \sectionskip\nobreak\noindent{\bf #1}\vskip0.3cm \noindent}

\nopagenumbers
\headline={\ifnum\pageno=\count31\frontheadline
  \else{\ifnum\pageno=0\frontheadline
     \else{{\raise 2pt\hbox to \hsize{\paperhead}}}\fi}\fi}

\footline={\centerline{\sevenbf \folio}}
\def\frontheadline{\sevenbf \hfil}
\def\paperhead{\sevenbf \heading\ \range\hfil\folio}
\newdimen\pagewidth \newdimen\pageheight \newdimen\ruleht
\maxdepth=2.2pt
\pagewidth=\hsize \pageheight=\vsize \ruleht=.5pt

\def\onepageout#1{\shipout\vbox{ 
    \offinterlineskip 
  \makeheadline
    \vbox to \pageheight{
         #1 
 \ifnum\pageno=\count31{\vskip 21pt\line{\the\footline}}\fi
 \ifvoid\footins\else 
 \vskip\skip\footins \kern-3pt
 \hrule height\ruleht width\pagewidth \kern-\ruleht \kern3pt
 \unvbox\footins\fi
 \boxmaxdepth=\maxdepth}
 \advancepageno}}
\output{\onepageout{\pagecontents}}
\count31=-1
\topskip 0.7 true in

\tolerance=10000
\centerline{\bf Nonsymmetric Gravitational Theory}
\centerline{\bf J. W. Moffat}
\centerline{\bf Department of Physics}
\centerline{\bf University of Toronto}
\centerline{\bf Toronto, Ontario M5S 1A7}
\centerline{\bf Canada}
\vskip 0.8 true in
\centerline{\bf Abstract}
\vskip 0.2 true in
A new version of nonsymmetric gravitational theory is presented.
The field equations are
expanded about the Minkowski metric, giving in lowest order the linear Einstein
field equations and massive Proca
field equations for the antisymmetric field $g_{[\mu\nu]}$.
An expansion about an arbitrary
Einstein background metric yields massive Proca field equations with couplings
to only physical modes. It follows that the new version of NGT
is free of ghost
poles, tachyons and higher-order poles and there are no problems with
asymptotic boundary conditions. A static spherically symmetric
solution of the field equations in the short-range approximation is everywhere
regular and does not contain a black hole event horizon.
\vskip 0.3 true in
UTPT-94-28. October, 1994.
\vskip 0.2 true in
e-mail: moffat@physics.utoronto.ca
\par\vfil\eject
\proclaim 1. {\bf Introduction}\par
\vskip 0.2 true in
A nonsymmetric gravitational theory (NGT), based on the decompositions
of the metric $g_{\mu\nu}$ and the connection $\Gamma^\lambda_{\mu\nu}$:
$$
g_{(\mu\nu)}={1\over 2}(g_{\mu\nu}+g_{\nu\mu}),\quad g_{[\mu\nu]}=
{1\over 2}(g_{\mu\nu}-g_{\nu\mu}),
\sectionit
$$
and
$$
\Gamma^\lambda_{\mu\nu}=\Gamma^\lambda_{(\mu\nu)}
+\Gamma^\lambda_{[\mu\nu]},
\sectionit
$$
is presented$^{1}$. The theory is free of ghosts, tachyons and
higher-order poles in the propagator in the
linear approximation. An expansion of $g_{\mu\nu}$ about an arbitrary
Einstein background metric also yields field equations to first order in
$g_{[\mu\nu]}$, which are free of couplings to unphysical (negative energy)
modes; the solutions of the field equations have consistent asymptotic boundary
conditions.

In view of the difficulty in obtaining physically consistent geometrical
generalizations of Einstein gravitational theory (EGT), it is interesting that
such a consistent theory can be formulated$^{2}$.

A static spherically symmetric solution of the NGT field equations in
the short-range approximation is everywhere regular and does not contain a
black
hole event horizon$^{3}$.
\vskip 0.2 true in
\setsection\proclaim 2. {\bf Nonsymmetric Gravitational Theory}\par
\vskip 0.2 true in
The non-Riemannian geometry is based on the nonsymmetric field structure with a
nonsymmetric $g_{\mu\nu}$ and affine connection $\Gamma^\lambda_{\mu\nu}$,
defined in Eqs.(1.1) and (1.2)$^{4}$.
The contravariant tensor $g^{\mu\nu}$ is defined in terms of the equation:
$$
g^{\mu\nu}g_{\sigma\nu}=g^{\nu\mu}g_{\nu\sigma}=\delta^\mu_\sigma.
\sectionit
$$

The Lagrangian density is given by
$$
{\cal L}_{NGT}={\cal L}_R+{\cal L}_M,
\sectionit
$$
where
$$
{\cal L}_R={\bf g}^{\mu\nu}R_{\mu\nu}(W)-2\lambda\sqrt{-g}
-{1\over 4}\mu^2{\bf g}^{\mu\nu}g_{[\nu\mu]}-{1\over 6}g^{\mu\nu}
W_\mu W_\nu,
\sectionit
$$
where $\lambda$ is the cosmological constant and $\mu^2$ is an
additional cosmological constant associated with $g_{[\mu\nu]}$.
Moreover, ${\cal L}_M$ is the matter Lagrangian density ($G=c=1$):
$$
{\cal L}_M=-8\pi g^{\mu\nu}{\bf T}_{\mu\nu}.
\sectionit
$$
Here, ${\bf g}^{\mu\nu}=\sqrt{-g}g^{\mu\nu}$ and $R_{\mu\nu}(W)$ is the
NGT contracted curvature tensor:
$$
R_{\mu\nu}(W)=W^\beta_{\mu\nu,\beta} - {1\over
2}(W^\beta_{\mu\beta,\nu}+W^\beta_{\nu\beta,\mu}) -
W^\beta_{\alpha\nu}W^\alpha_{\mu\beta} +
W^\beta_{\alpha\beta}W^\alpha_{\mu\nu},
\sectionit
$$
defined in terms of the unconstrained nonsymmetric connection:
$$
W^\lambda_{\mu\nu}=\Gamma^\lambda_{\mu\nu}-{2\over 3}\delta^\lambda_\mu
W_\nu,
\sectionit
$$
where
$$
W_\mu={1\over 2}(W^\lambda_{\mu\lambda}-W^\lambda_{\lambda\mu}).
\sectionit
$$
Eq.(2.6) leads to the result:
$$
\Gamma_\mu=\Gamma^\lambda_{[\mu\lambda]}=0.
\sectionit
$$

The NGT contracted curvature tensor can be written as
$$
R_{\mu\nu}(W) = R_{\mu\nu}(\Gamma) + {2\over 3} W_{[\mu,\nu]},
\sectionit
$$
where $R_{\mu\nu}(\Gamma)$ is defined by
$$
R_{\mu\nu}(\Gamma ) = \Gamma^\beta_{\mu\nu,\beta} -{1\over 2}
\left(\Gamma^\beta_{(\mu\beta),\nu} + \Gamma^\beta_{(\nu\beta),\mu}\right) -
\Gamma^\beta_{\alpha\nu} \Gamma^\alpha_{\mu\beta} +
\Gamma^\beta_{(\alpha\beta)}\Gamma^\alpha_{\mu\nu}.
\sectionit
$$

The field equations in the presence of matter sources are given by:
$$
G_{\mu\nu} (W)+\lambda g_{\mu\nu}+{1\over 4}\mu^2C_{\mu\nu}
-{1\over 6}(P_{\mu\nu}-{1\over 2}g_{\mu\nu}P) = 8\pi T_{\mu\nu},
\sectionit
$$
$$
{{\bf g}^{[\mu\nu]}}_{,\nu} = -{1\over 2}{\bf g}^{(\mu\beta)}W_\beta,
\sectionit
$$
$$
{{\bf g}^{\mu\nu}}_{,\sigma}+{\bf g}^{\rho\nu}W^\mu_{\rho\sigma}
+{\bf g}^{\mu\rho}
W^\nu_{\sigma\rho}-{\bf g}^{\mu\nu}W^\rho_{\sigma\rho}
+{2\over 3}\delta^\nu_\sigma{\bf g}^{\mu\rho}W^\beta_{[\rho\beta]}
$$
$$
+{1\over 6}({\bf g}^{(\mu\beta)}W_\beta\delta^\nu_\sigma
-{\bf g}^{(\nu\beta)}W_\beta\delta^\mu_\sigma)=0.
\sectionit
$$
Here, we have
$$
G_{\mu\nu} = R_{\mu\nu} - {1\over 2} g_{\mu\nu} R,
\sectionit
$$
$$
C_{\mu\nu}=g_{[\mu\nu]}+{1\over 2}g_{\mu\nu}g^{[\sigma\rho]}
g_{[\rho\sigma]}+g^{[\sigma\rho]}g_{\mu\sigma}g_{\rho\nu},
\sectionit
$$
$$
P_{\mu\nu}=W_\mu W_\nu,
\sectionit
$$
and $P=g^{\mu\nu}P_{\mu\nu}=g^{(\mu\nu)}W_\mu W_\nu$.

In empty space, the field equations (2.11) become:
$$
R_{\mu\nu}(\Gamma)={2\over 3}W_{[\nu,\mu]}+\lambda
g_{\mu\nu}-{1\over 4}\mu^2(C_{\mu\nu}-{1\over 2}g_{\mu\nu}C)
+{1\over 6}P_{\mu\nu}.
\sectionit
$$

The generalized Bianchi identities:
$$
[{\bf g}^{\alpha\nu}G_{\rho\nu}(\Gamma)+{\bf g}^{\nu\alpha}
G_{\nu\rho}(\Gamma)]_{,\alpha}+{g^{\mu\nu}}_{,\rho}{\bf G}_{\mu\nu}=0,
\sectionit
$$
give rise to the matter response equations:
$$
g_{\mu\rho}{{\bf T}^{\mu\nu}}_{,\nu}+g_{\rho\mu}{{\bf T}^{\nu\mu}}_{,\nu}
+(g_{\mu\rho,\nu}+g_{\rho\nu,\mu}-g_{\mu\nu,\rho}){\bf T}^{\mu\nu}=0.
\sectionit
$$

In general, the last term in Eq.(2.3) is of the form, ${1\over 2}\sigma
g^{\mu\nu}W_\mu W_\nu$. The coupling constant $\sigma$ is chosen to
be $\sigma=-{1\over 3}$, so that the theory yields consistent ghost
and tachyon free perturbative solutions to the field equations.
\vskip 0.2 true in
\setsection\proclaim 3. {\bf Linear Approximation}\par
\vskip 0.2 true in
Let us assume that $\lambda=0$ and expand $g_{\mu\nu}$ about Minkowski
spacetime:
$$
g_{\mu\nu}=\eta_{\mu\nu}+{}^{(1)}h_{\mu\nu}+...,
\sectionit
$$
where $\eta_{\mu\nu}$ is the Minkowski metric tensor: $\eta_{\mu\nu}=
\hbox{diag}(-1, -1, -1, +1)$. We also expand $\Gamma^\lambda_{\mu\nu}$ and
$W^\lambda_{\mu\nu}$:
$$
\Gamma^\lambda_{\mu\nu}={}^{(1)}\Gamma^\lambda_{\mu\nu}
+{}^{(2)}\Gamma^\lambda_{\mu\nu}+...,
$$
$$
W_\mu={}^{(1)}W_\mu+{}^{(2)}W_\mu+...\,.
\sectionit
$$

Let us adopt the notation: $\psi_{\mu\nu}={}^{(1)}h_{[\mu\nu]}$. To first order
of approximation, Eq.(2.12) gives
$$
\psi_\mu=-{1\over 2}W_\mu,
\sectionit
$$
where for convenience $W_\mu$ denotes ${}^{(1)}W_\mu$. Moreover,
$$
\psi_\mu={\psi_{\mu\beta}}^{,\beta}=\eta^{\beta\sigma}\psi_{\mu\beta,\sigma}.
\sectionit
$$
{}From Eq.(2.13), we obtain the result to first order:
$$
{}^{(1)}{\Gamma^\lambda_{\mu\nu}}={1\over 2}\eta^{\lambda\sigma}
({}^{(1)}h_{\sigma\nu,\mu}
+{}^{(1)}h_{\mu\sigma,\nu}-{}^{(1)}h_{\nu\mu,\sigma})+{1\over 6}
({\delta^\lambda_\nu}W_\mu-{\delta^\lambda_\mu}W_\nu).
\sectionit
$$

The antisymmetric and symmetric field equations derived from Eq.(2.11)
decouple to lowest order; the symmetric equations are the usual
Einstein field equations in the linear approximation. The skew
equations are given by
$$
(\dal+\mu^2)\psi_{\mu\nu}=J_{\mu\nu},
\sectionit
$$
where
$$
J_{\mu\nu}=16\pi(T_{[\mu\nu]}+{2\over \mu^2}{T_{[[\mu\sigma],\nu]}}^{,\sigma}).
\sectionit
$$
We have
$$
W_\mu=-{32\pi\over \mu^2}{T_{[\mu\nu]}}^{,\nu},
\sectionit
$$
and from Eq.(3.3) we get
$$
\psi_\mu={16\pi\over \mu^2}{T_{[\mu\nu]}}^{,\nu}.
\sectionit
$$

In the wave-zone, $T_{\mu\nu}=0$, and Eqs.(3.6) and (3.9) become
$$
(\dal+\mu^2)\psi_{\mu\nu}=0,
\sectionit
$$
$$
\psi_\mu=0.
\sectionit
$$
These equations can be obtained from the Lagrangian:
$$
{\cal L}_{\psi}={1\over 4}\psi^2_{\mu\nu,\lambda}-{1\over 2}\psi_\mu^2
-{1\over 4}\mu^2\psi_{\mu\nu}^2.
\sectionit
$$
The $\psi_{\mu\nu}$ has the spin decomposition:
$$
\psi_{\mu\nu}=1_b\oplus 1_e,
\sectionit
$$
where $1_b$ and $1_e$ denote the magnetic and electric vectors, respectively.
Only the magnetic vector propagates corresponding to a massive spin
$1^+$ pseudovector field with the propagator:
$$
\Pi={P^1_b\over k^2},
\sectionit
$$
where $P^1_b$ is the magnetic projection operator defined by
$$
P^1_b={1\over 2}(\theta_{\mu\rho}\theta_{\nu\sigma}-\theta_{\mu\sigma}
\theta_{\nu\rho}),
\sectionit
$$
$$
\theta_{\mu\nu}=\delta_{\mu\nu}-{k_\mu k_\nu\over k^2},
\sectionit
$$
and $k_\mu$ denotes the momentum four vector. The Lagrangian (3.12) is free of
ghosts, tachyons and higher-order poles$^{5}$ and the Hamiltonian is positive
and bounded from below.

The physical
requirement that the weak field linear approximation of the NGT field
equations is free of ghosts, tachyons and higher-order poles in the
propagator is satisfied by the version of NGT presented here. This overcomes
the problem of obtaining a consistent geometrical NGT, in which the
physical vacuum in the theory is stable.

As shown by van Nieuwenhuizen$^{5}$, in flat Minkowski spacetime there
exist only two physically consistent
Lagrangians for the antisymmetric potential field $\psi_{\mu\nu}$, which
are free of ghosts, tachyons and higher-order poles in
the propagator, namely, the massless gauge invariant
theory, and the massive Proca-type theory with the Lagrangian (3.12).
Since NGT does not possess a massless gauge invariance, then the only other
possibility is that it should reduce to the physical, massive Proca-type
theory. We have, in fact, now discovered a fully geometrical NGT scheme
that fulfils this requirement.
\par\vfil\eject
\setsection\proclaim 4. {\bf Expansion of the Field Equations Around a Curved
Background}\par
\vskip 0.2 true in
Let us now consider the expansion of the field equations around an arbitrary
Einstein background metric. We shall introduce the notation: $g_{[\mu\nu]}
=a_{\mu\nu}$. We have
$$
g_{\mu\nu}=g_{S\mu\nu}+{}^{(1)}g_{\mu\nu}+...,\quad \Gamma^\lambda_{\mu\nu}
=\Gamma^\lambda_{S\mu\nu}+{}^{(1)}{\Gamma^\lambda_{\mu\nu}}+...,\quad
W_\mu={}^{(1)}W_\mu+{}^{(2)}W_\mu+...,
\sectionit
$$
where $g_{S\mu\nu}$ and $\Gamma^\lambda_{S\mu\nu}$ denote the Einstein
background metric tensor and connection, respectively, and
$g_S^{\lambda\alpha}g_{S\beta\lambda}=\delta^\alpha_\beta$. We also
introduce the notation for the Riemann tensor:
$$
{B^\sigma}_{\mu\nu\rho}=\Gamma^\sigma_{S\mu\nu,\rho}
-\Gamma^\sigma_{S\mu\rho,\nu}
-\Gamma^\sigma_{S\alpha\nu}\Gamma^\alpha_{S\mu\rho}
+\Gamma^\sigma_{S\alpha\rho}\Gamma^\alpha_{S\mu\nu}.
\sectionit
$$
By performing the contraction on the suffixes $\sigma$ and $\rho$,
we get the Ricci tensor, $B_{\mu\nu}={B^\alpha}_{\mu\nu\alpha}$.
We get to first order in $a_{\mu\nu}$ and $W_\mu$ (we denote
${}^{(1)}W_\mu$ by $W_\mu$ and ${}^{(1)}a_{\mu\nu}$ by $a_{\mu\nu}$) the field
equations:
$$
B_{\mu\nu}=0,
\sectionit
$$
$$
\nabla^\sigma a_{\mu\sigma}={1\over 2\mu^2}
\biggl[\nabla^\nu\biggl(4g_S^{\lambda\sigma}g_S^{\alpha\beta}
B_{\alpha\nu\lambda\mu}a_{\sigma\beta}
-2(Ba)_{\mu\nu}\biggr)\biggr].
\sectionit
$$
$$
(\dal_S+\mu^2)a_{\mu\nu}=M_{\mu\nu},
\sectionit
$$
where
$$
M_{\mu\nu}=2g_S^{\lambda\sigma}g_S^{\alpha\beta}
B_{\alpha\nu\lambda\mu}a_{\sigma\beta}
+{1\over \mu^2}\nabla_{[\nu}\nabla^\rho\biggl[4g_S^{\lambda\sigma}
g_S^{\alpha\beta}B_{\alpha\rho\lambda\mu]}a_{\sigma\beta}
-2(Ba)_{\mu]\rho}\biggr],
\sectionit
$$
and $\dal_S=\nabla^\sigma\nabla_\sigma,\nabla^\sigma=g_S^{\sigma\alpha}
\nabla_\alpha$. Also, $(Ba)_{\mu\nu}$ denotes additional terms involving
products of the Riemann tensor and $a_{\mu\nu}$.

As before, $W_\mu$ does not propagate and there is no coupling to unphysical
modes through the effective source tensor formed from the Riemann tensor and
$a_{\mu\nu}$.

The energy associated with the flux of gravitational waves, calculated in the
wave-zone for $r\rightarrow \infty$, is positive definite.
\vskip 0.2 true in
\setsection\proclaim 5.
{\bf Static Spherically Symmetric Solution}
\par\vskip 0.2 true in
In the case of a static spherically symmetric field,
Papapetrou has derived the canonical form of $g_{\mu\nu}$ in NGT$^{6}$:
$$
g_{\mu\nu}=\left(\matrix{-\alpha&0&0&w\cr
0&-\beta&f\hbox{sin}\theta&0\cr 0&-f\hbox{sin}\theta&
-\beta\hbox{sin}^2
\theta&0\cr-w&0&0&\gamma\cr}\right),
\sectionit
$$
where $\alpha,\beta,\gamma$ and $w$ are functions of $r$.
For the theory in which there is no NGT magnetic monopole charge, we have
$w=0$ and only the $g_{[23]}$ component of $g_{[\mu\nu]}$ survives.

The line element for a spherically symmetric body is given by
$$
ds^2=\gamma(r)dt^2-\alpha(r)dr^2-\beta(r)(d\theta^2+\hbox{sin}^2\theta
d\phi^2).
\sectionit
$$
We have
$$
\sqrt{-g}=\hbox{sin}\theta[\alpha\gamma(\beta^2+f^2)]^{1/2}.
\sectionit
$$

The vector $W_\mu$ can be determined from Eq.(2.12):
$$
W_\mu=-2k_{\mu\rho}{{\bf g}^{[\rho\sigma]}}_{,\sigma},
\sectionit
$$
where $k_{\mu\nu}$ is defined by $k_{\mu\alpha}g^{(\mu\beta)}
=\delta_\alpha^\beta$. For the static spherically symmetric field with
$w=0$ it follows from (5.4) that $W_\mu=0$.

We choose $\lambda=0, \beta=r^2$ and demand the boundary conditions:
$$
\alpha\rightarrow 1,\quad \gamma\rightarrow 1,\quad f\rightarrow 0,
\sectionit
$$
as $r\rightarrow \infty$. We also assume the short-range approximation for
which the $\mu^2$ contributions in the vacuum field equations can be neglected.
If $\mu^{-1} >> 2m$, then we can use the static, spherically symmetric Wyman
solution$^{7}$:
$$
\gamma=\hbox{exp}(\nu),
\sectionit
$$
$$
\alpha=m^2(\nu^\prime)^2\hbox{exp}(-\nu)(1+s^2)
(\hbox{cosh}(a\nu)-\hbox{cos}(b\nu))^{-2},
\sectionit
$$
$$
f=[2m^2\hbox{exp}(-\nu)(\hbox{sinh}(a\nu)\hbox{sin}(b\nu)
$$
$$
+s(1-\hbox{cosh}(a\nu)\hbox{cos}(b\nu))]
(\hbox{cosh}(a\nu)-\hbox{cos}(b\nu))^{-2},
\sectionit
$$
where $\nu$ is implicitly determined by the equation:
$$
\hbox{exp}(\nu)(\hbox{cosh}(a\nu)-\hbox{cos}(b\nu))^2{r^2\over 2m^2}=
\hbox{cosh}(a\nu)\hbox{cos}(b\nu)-1+s\hbox{sinh}(a\nu)\hbox{sin}(b\nu).
\sectionit
$$
Here, $s$ is a dimensionless constant of integration.

We find for $2m/r < 1$ and $0 < sm^2/r^2 < 1$ that $\alpha$ and $\gamma$
take the Schwarzschild form:
$$
\gamma=\alpha^{-1}=1-{2m\over r}.
\sectionit
$$

Near $r=0$ we can develop expansions where $r/m < 1$ and $0 < \vert s\vert
<1$.
The leading terms are
$$
\gamma=\gamma_0+{\gamma_0(1+{\cal O}(s^2))\over 2\vert s\vert}\biggl({r\over
m}\biggr)^2 + {\cal O}\biggl(\biggl({r\over m}\biggr)^4\biggr),
\sectionit
$$
$$
\alpha={4\gamma_0(1+{\cal O}(s^2))\over s^2}\biggl({r\over m}\biggr)^2
+{\cal O}\biggl(\biggl({r\over m}\biggr)^4\biggr),
\sectionit
$$
$$
f=m^2\biggl(4-{\vert s\vert\pi\over 2}+s\vert s\vert+{\cal O}(s^3)\biggr)
+{\vert s\vert+s^2\pi/8+{\cal O}(s^3)\over 4}r^2+{\cal O}(r^4),
\sectionit
$$
$$
\gamma_0=\hbox{exp}\biggl(-{\pi+2s\over \vert s\vert}+{\cal O}(s)\biggr)...\,.
\sectionit
$$
These solutions clearly illustrate
the non-analytic nature of the limit $s\rightarrow 0$ in the strong
gravitational field regime.

The singularity caused by the vanishing of $\alpha(r)$ at
$r=0$ is a {\it coordinate} singularity, which can be removed by
transforming to another coordinate frame of reference$^{3}$. The curvature
invariants do not, of course, contain any coordinate singularities.

The NGT curvature invariants such as the generalized Kretschmann scalar:
$$
K=R^{\lambda\mu\nu\rho}R_{\lambda\mu\nu\rho}
\sectionit
$$
are finite.

The solution is everywhere non-singular and
there is no event horizon at $r=2m$. A black hole is replaced in the theory
by a superdense object which can be stable for an arbitrarily large
mass$^{3,8}$.
\vskip 0.2 true in
\setsection\proclaim 6.
{\bf Conclusions}\par
\vskip 0.2 true in
We have succeeded in deriving a geometrical version of NGT
which yields a stable, ghost and tachyon free linear approximation, when
the field equations are expanded about Minkowski spacetime or about an
arbitrary Einstein background metric.

In the short-range approximation, in which $\mu^{-1}$ is large compared to
$2m$, a static spherically symmetric
solution can be obtained from the field equations which is regular everywhere
in spacetime and which does not contain an event horizon at $r=2m$. The
black hole that is predicted by the collapse of a sufficiently massive
star in EGT is replaced by a superdense object that is stable for an
arbitrarily large mass.

An important result of this study is that a classical
theory of gravity can be formulated, which has a physically consistent
perturbative expansion for weak fields, and does not have singularities and
black holes. It satisfies the standard gravitational experimental tests.
Recently, a solution has been published for inhomogeneous gravitational
collapse of a matter cloud with a general form of matter, which leads
to a naked singularity$^{9}$. The collapse is related to the choice of
initial data for the Einstein field equations, and the naked singularity
would occur in generic situations based on regular initial conditions
satisfying the weak energy condition. This result would lead to the
demise of EGT, for it represents a local violation of the Cauchy data for
collapse; it would provide a strong motivation for seriously considering
a classical gravity theory such as NGT with everywhere regular solutions
of the field equations.

Because there is no event horizon in the static spherically symmetric solution,
we can resolve the problem of information loss$^{10}$ associated with black
holes at a classical level.
\vskip 0.2 true in
{\bf Acknowledgements}
\vskip 0.2 true in
I thank M. Clayton, N. J. Cornish, J. L\'egar\'e, P. Savaria and I. Yu.Sokolov
for helpful and stimulating
discussions. I thank the Natural Sciences and Engineering Research Council of
Canada for the support of this work.
\vskip 0.1 true in
\centerline{\bf References}
\vskip 0.1 true in
\item{1.}{For earlier work on NGT, see: J. W. Moffat, Phys. Rev. D {\bf 19},
3554 (1979); J. W. Moffat, J. Math. Phys. {\bf 21}, 1978 (1980); J. W. Moffat,
Phys. Rev. D {\bf 35}, 3733 (1987). For a review of NGT, see: J. W. Moffat,
Proceedings of the Banff Summer Institute on Gravitation, Banff, Alberta,
August 12-25, 1990, edited by R. Mann and P. Wesson, World Scientific, p.523,
1991.}
\item{2.}{R. B. Mann and J. W. Moffat, J. Phys. A {\bf 14}, 2367 (1981);
T. Damour, S. Deser and J. McCarthy, Phys. Rev. D {\bf 47}, 1541
(1993).}
\item{3.}{N. J. Cornish and J. W. Moffat, Phys. Letts. {\bf 336B}, 337 (1994);
University of Toronto preprint, UTPT-94-08, 1994, gr-qc/9406007. To be
published
in J. Math. Phys.; I. Yu.Sokolov, University of Toronto preprint, UTPT-94-23,
1994.}
\item{4.}{A. Einstein, The Meaning of Relativity, Menthuen and Co. London,
U.K. fifth edition, Appendix II, 1951. Einstein used the nonsymmetric field
structure to derive a unified field theory of gravitation and
electromagnetism. The nonsymmetric field equations cannot produce
a consistent physical theory of electromagnetism. We use the nonsymmetric
field structure to descibe a non-Riemannian theory of gravitation.}
\item{5.}{P. van Nieuwenhuizen, Nucl. Phys. B {\bf 60}, 478 (1973).}
\item{6.}{A. Papapetrou, Proc. Roy. Ir. Acad. Sec. A {\bf 52}, 69 (1948).}
\item{7.}{M. Wyman, Can. J. Math. {\bf 2}, 427 (1950).}
\item{8.}{N. J. Cornish, University of Toronto preprint, UTPT-94-10, 1994.
Revised November, 1994.}
\item{9.}{P. S. Joshi and I. H. Dwivedi, Commun. Math. Phys. {\bf 146},
333 (1992); to appear in Commun. Math. Phys. (1994); P. S. Joshi, Global
Aspects in Gravitation and Cosmology, Clarendon Press, Oxford, 1994.}
\item{10.}{S. Hawking, Phys. Rev. D {\bf 14}, 2460 (1976); Commun. Math.
Phys. {\bf 87}, 395 (1982).}

\end